\documentclass[a4paper]{jpconf}
\usepackage{graphicx}
\begin{document}
\title{Magnetic field dynamos and magnetically triggered flow instabilities}

\author{F Stefani$^1$, T Albrecht$^2$, R Arlt$^3$, M Christen$^4$,
A Gailitis$^5$, 
M Gellert$^3$, 
A Giesecke$^1$, O Goepfert$^6$, J Herault$^7$, O N Kirillov$^8$, 
G. Mamatsashvili$^1$,
J Priede$^9$, G R\"udiger$^3$, M Seilmayer$^1$, A Tilgner$^6$  and T Vogt$^1$}

\address{$^1$ Helmholtz-Zentrum Dresden -- Rossendorf,
Bautzner Landstra\ss e 400, D-01328 Dresden, Germany}
\address{$^2$ Department of Mechanical and Aerospace Engineering, Monash University,
VIC 3800, Australia}
\address{$^3$ Leibniz-Institut f\"ur Astrophysik Potsdam, 
An der Sternwarte 16, D-14482 Potsdam, Germany}
\address{$^4$ Technische Universit\"at Dresden, Institut f\"ur Energietechnik,
D-01062 Dresden, Germany}
\address{$^5$ Institute of Physics, University of Latvia, LV-2169 Salaspils, Miera iela 32, Latvia}
\address{$^6$ Institute of Geophysics, University of G\"ottingen, Friedrich-Hund-Platz 1, 
D-37077 G\"ottingen, Germany}
\address{$^7$ Laboratoire des Sciences du Num{\'e}rique de Nantes (LS2N),
CNRS, IMT Atlantique, Nantes, France}
\address{$^8$ Northumbria University,
Mathematics, Physics and Electrical Engineering, Ellison Building, D219
Newcastle upon Tyne NE1 8ST, United Kingdom}
\address{$^9$ Flow Measurement Research Centre, Coventry University, UK}

\ead{F.Stefani@hzdr.de}

\begin{abstract}
The project A2 of the LIMTECH Alliance aimed at a better  
understanding of those magnetohydrodynamic 
instabilities that are relevant for the generation 
and the action of cosmic magnetic fields. These comprise 
the hydromagnetic dynamo effect and 
various magnetically triggered flow instabilities, such as the 
magnetorotational instability and the Tayler instability. 
The project was intended to support the experimental capabilities 
to become available in the framework of the DREsden Sodium 
facility for DYNamo and thermohydraulic studies (DRESDYN).
An associated starting grant was focused on the dimensioning of 
a liquid metal experiment on the newly found 
magnetic destabilization of rotating flows with positive 
shear. In this paper, the main results of these
two projects are summarized.

\end{abstract}

\section{Introduction}

Magnetic fields of planets, stars and galaxies are generated 
by the homogeneous dynamo effect 
\cite{Jones_2011,Charbonneau_2014,Beck_2009}. Once produced,
cosmic magnetic fields can play a surprisingly active role 
in cosmic structure formation via 
various magnetically triggered flow instabilities,
such as the celebrated magnetorotational instability (MRI) 
\cite{Balbus_2003} and the current-driven  Tayler instability 
(TI) \cite{Tayler_1973}.

Complementary to the decades-long theoretical and numerical
efforts to understand these fundamental magnetohydrodynamic 
effects, the last years have 
seen great progress in dedicated experimental investigations
\cite{Gailitis_2002,Stefani_2008,Lathrop_2011}. 
After the pioneering Riga and Karlsruhe dynamo experiments
\cite{Gailitis_2000,Gailitis_2001,Stieglitz_2001}, it was in particular 
the rich dynamics observed in the French von K\'arm\'an Sodium (VKS) 
experiment \cite{Berhanu_2010} that provoked much interest 
throughout the dynamo community.
The observed reversals, excursions, bursts, 
hemispherical fields etc. inspired new activities to understand 
the essential physics behind the corresponding planetary 
phenomena \cite{Stefani_2006a,Sorriso_2007,Petrelis_2009,Benzi_2010}.

For magnetically triggered flow instabilities, 
the situation is more subtle.
Interesting results had been obtained in 
a liquid sodium spherical Couette experiment in Maryland 
in form of coherent velocity/magnetic field fluctuations 
showing up in a parameter region reminiscent of MRI \cite{Sisan_2004}, 
as well as in the 
GaInSn Taylor-Couette (TC) experiment in Princeton which 
provided evidence for slow magneto-Coriolis waves 
\cite{Nornberg_2010} and a free-Shercliff layer 
instability \cite{Roach_2012}.
Despite these achievements, both experiments have 
corroborated the intricacies of demonstrating the standard 
version of MRI (SMRI) for which a purely axial field is applied. 
For liquid metal TC flows, in particular, the complications result 
from the compromising effect of axial boundaries
on the flow structure at those high Reynolds numbers 
($>10^6$) that are necessary 
if magnetic Reynolds numbers of order 10  
are required.
 
More conclusive, albeit less ambitious, 
was the experimental demonstration
of two special types of the MRI which arise
when applying helical or purely azimuthal magnetic 
fields to the rotating flow. These instabilities 
have been coined helical MRI (HMRI)
and azimuthal MRI (AMRI), respectively.
As was first shown by Hollerbach and R\"udiger
\cite{Hollerbach_2005}, the essentially 
inductionless, axisymmetric ($m=0$) HMRI scales with the
Reynolds and Hartmann number rather than
with magnetic Reynolds and 
Lundquist number as SMRI with which it 
is monotonically connected, though 
\cite{Hollerbach_2005,Kirillov_2010}. 
It is this scaling behaviour which makes 
experimental investigations of HMRI much easier than those
of SMRI.
Indeed,  first  evidence of HMRI occurring 
in the predicted parameter regions 
of Hartmann number  with roughly correct eigenfrequencies
was demonstrated  in the
PROMISE experiment at Helmholtz-Zentrum Dresden-Rossendorf (HZDR)
\cite{Stefani_2006b,Stefani_2007}.
In 2009, an improved version of this 
experiment - using split end-rings installed 
at the top and bottom of the cylinder in order 
to minimize the global effects of 
Ekman pumping - allowed to characterize
HMRI by a number of parameter variations, generally 
in good agreement with numerical predictions 
\cite{Stefani_2009a}. 

Nearly at the same time, Hollerbach et al. 
\cite{Hollerbach_2010} identified AMRI 
as a second induction-less 
MRI version that appears for strongly 
dominant azimuthal fields in form of a 
non-axisymmetric ($m=1$) perturbation.
When relaxing the condition, implicit 
for AMRI and HMRI,  
that the azimuthal field should be 
current-free in the liquid, one enters the 
vast field of current-driven instabilities
which 
includes the Tayler instability (TI) 
\cite{Tayler_1973}. This kink-type 
instability, whose ideal counterpart has long been 
known in plasma physics \cite{Bergerson_2006}, 
is also discussed as a central 
mechanism of the non-linear 
Tayler-Spruit dynamo model for stellar magnetic fields 
\cite{Spruit_2002,Gellert_2008,Ruediger_2011}).
While TI was experimentally observed
prior to the start of the LIMTECH alliance 
\cite{Seilmayer_2012}, AMRI was first demonstrated  
within the funding period
\cite{Seilmayer_2014}.

Despite -- and partly inspired by -- those experimental 
achievements, there are 
still a number of questions worth to be studied in
laboratory.
While realistic ''bonsai'' models of cosmic objects, 
with all dimensionless numbers matching those of planets
or stars, 
are certainly outside the scope of laboratory feasibility 
\cite{Lathrop_2011}, some new facilities still 
challenge the physical and
technical limits of dynamo experiments. 
This certainly applies to the 3\,m diameter spherical Couette experiment 
running at the University of Maryland 
\cite{Zimmermann_2010,Adams_2015}, 
to the 3\,m diameter 
plasma dynamo experiment in Madison
\cite{Cooper_2014,Weisberg_2017}, as well as
to the 2\,m diameter precession dynamo experiment 
presently under construction at HZDR
\cite{Stefani_2012,Stefani_2015}. One of the 
unwelcome characteristics of these ''second generation'' 
dynamo experiments is a higher uncertainty 
of success. The Riga and Karlsruhe experiments
were quite accurately described by kinematic 
dynamo codes and 
simplified saturation models 
\cite{Gailitis_2008,Tilgner_2001,Raedler_2003}, 
and even the unexpected VKS dynamo results were understood once 
the effect of the high permeability disks 
was accounted for properly
\cite{Giesecke_2010,Giesecke_2012,Nore_2015,Nore_2016}.
By contrast,  
the outcomes of the 
experiments in Maryland, Madison, and Dresden 
are much harder to predict. 
In either case, this uncertainty results from the 
ambition to construct a truly 
homogeneous dynamo, which is neither driven by pumps or propellers, 
nor influenced by guiding blades or gradients of magnetic 
permeability. This higher degree of freedom makes those 
flows prone to exhibiting medium-size flow structures 
and waves, whose dynamo capabilities are still under scrutiny. 

On the MRI side, the ''holy grail'' of observing
SMRI in the lab is yet to be found. One promising
set-up is the relatively 
flat TC experiment in Princeton
which circumvents the Ekman-pumping induced 
distortion of  the original TC flow profile by 
separately driving different rings of the lids.
Another promising way is followed in the plasma experiment in 
Madison where already reasonable
radial velocity profiles have been produced by a near-wall 
${\bf j} \times {\bf B}$ 
driving of argon and helium plasmas \cite{Weisberg_2017}.
At HZDR, a more traditional path towards SMRI is 
pursued: the respective experiment contains a long  
sodium column between two rotating cylinders 
exposed to a strong 
axial magnetic field. Yet, this axial field
is complemented by an  azimuthal magnetic field,
which permits progressing from the well-known
regime of HMRI towards the limit of
SMRI by increasing the Reynolds and 
Hartmann number and simultaneously decreasing the ratio of
azimuthal to axial field.

\begin{figure}[ht]
\begin{center}
\includegraphics[width=0.85\textwidth]{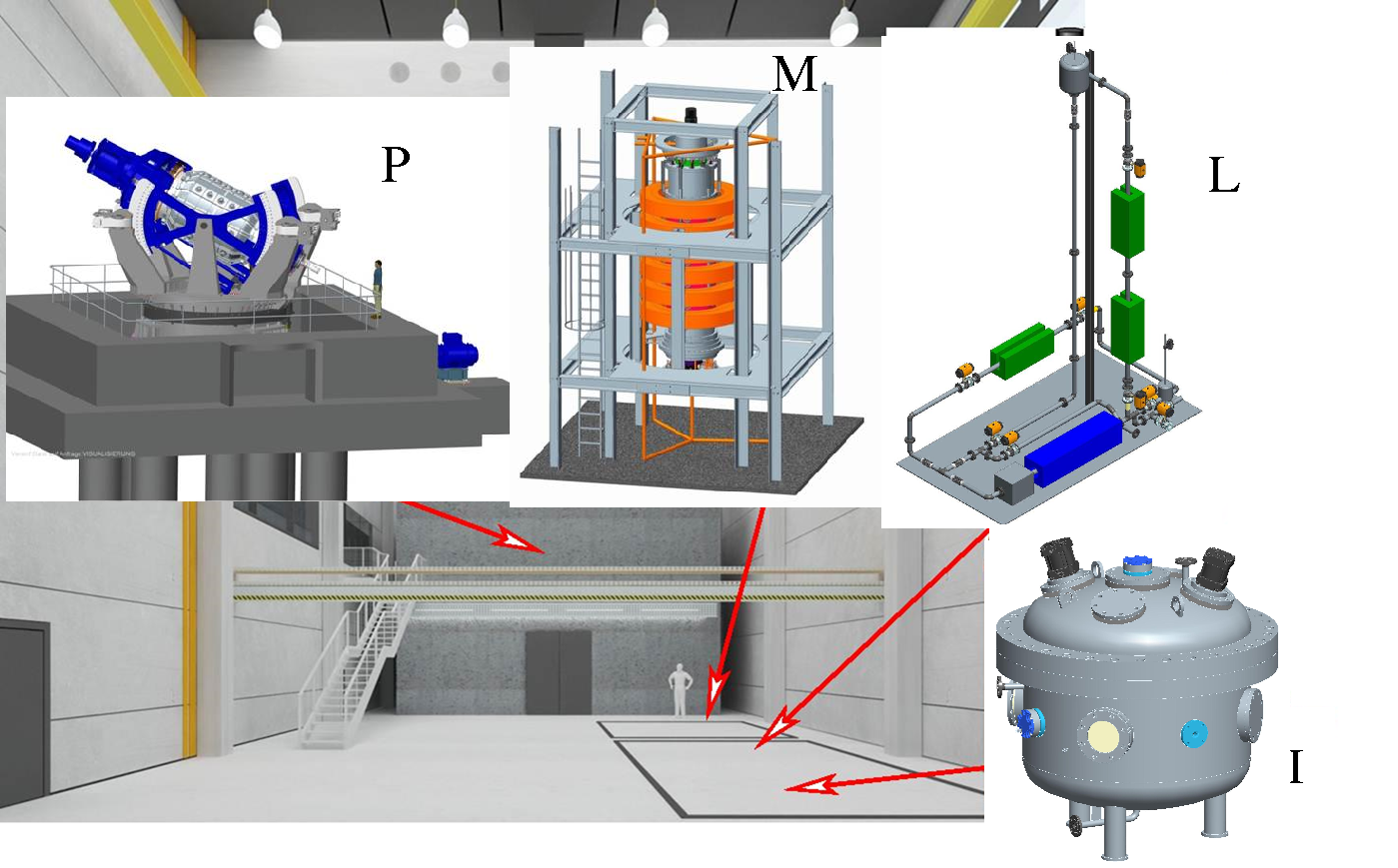}
\end{center}
\caption{Interior of the central hall of the DRESDYN facility
with the main planned experiments. Precession driven 
dynamo experiment (P) to be 
installed in the containment; Tayler-Couette experiment 
for the investigation of the magnetorotational and the 
Tayler instability (M); sodium loop (L); In-Service-Inspection
experiment (I). A further test stand for liquid 
metal batteries is also foreseen but not yet designed.}
\label{fig:dresdyn}
\end{figure}

This paper summarizes research activities within the 
LIMTECH Alliance that were dedicated to preparations of 
the precession-driven dynamo experiment and  the large-scale
MRI experiment. Both experiments are presently 
under construction in frame of the DRESDYN project 
at HZDR, which will include also a number of other 
experiments with liquid sodium 
(see Figure \ref{fig:dresdyn}).
On the dynamo side, activities included a number of numerical
simulations of precession driven flows and their dynamo action 
in cylinders and cubes, as well as experimental work at a 1:6 
down-scaled water experiment.
They also comprised various activities to 
refurbish and recommission the 
Riga dynamo experiment which -- apart from
having its own scientific goals -- serves also for
testing various measurement techniques for 
future DRESDYN experiments.

With regard to  MRI/TI, the project supported 
the first demonstration of AMRI at the PROMISE facility
at HZDR, and included
theoretical and numerical work on
various aspects of 
the interaction of rotating fluids and magnetic fields.
These led to the characterization of a new
magnetically triggered instability (''Super-AMRI'') 
that destabilizes rotating flows with
positive shear,  to the ''refutation'' of 
Chandrasekhar's theorem for magnetized rotating flows, 
and to the establishment of a rigorous
mathematical connection between the non-modal growth
for purely hydrodynamic rotating flows and
the growth rate of helical MRI.

\section{Dynamos}

In this section we summarize the main activities dedicated to 
existing and planned dynamo experiments. While 
the main focus was on various preparations of the DRESDYN 
precession driven dynamo, we will start with presenting 
the works 
related to the re-commissioning of the Riga dynamo.

\subsection{Riga dynamo experiment}
On 11 November 1999 the kinematic phase of
magnetic-field self-excitation was shortly 
observed at the Riga 
dynamo facility  before the experiment 
had to be stopped due to a minor leakage of 
liquid sodium  \cite{Gailitis_2000}.
After some repairs, in July 2000, a number of 
full runs clearly demonstrated both the kinematic and
the saturated dynamo regime \cite{Gailitis_2001}, thereby 
laying the basis for a comprehensive 
data base including  growth rates, frequencies, and 
spatial structures of 
the magnetic eigenfield in dependence on the impeller's
rotation rate \cite{Gailitis_2008}. 
Data of the kinematic regime was shown to 
be in very good agreement with
numerical predictions \cite{Stefani_1999}, 
and even the saturation regime was 
reasonably understood by applying Lenz's rule to
the specifics of this hydrodynamic dynamo 
\cite{Gailitis_2004}.

After a series of experimental campaigns, 
which delivered quite reliable und reproducible results,
the Riga dynamo was disassembled in order to replace 
and stabilize an inner cylinder that had been deformed 
during one run in 2010. The project
A2 supported the refurbishment with the 
double aim of testing measurement techniques for the 
DRESDYN precession dynamo, and of
preparing such a modification of the experiment 
that allows for observing new and 
non-trivial back-reaction effects.

The latter was motivated by the numerical finding  that
a specific ''de-optimization'' of the flow
field in the Riga dynamo could lead to
a vacillation between two different states 
of the dynamo \cite{Stefani_2011}. Actually, we set out
from the hypothesis  that a too high 
initial azimuthal component of the
velocity might yield a subcritical Hopf 
bifurcation, just by virtue of a selective breaking 
of this component that provides  a ''re-optimization'' 
of the velocity. Instead of such a ''hard'' 
subcritical Hopf bifurcation we found numerically
a ''soft'' vacillation between two dynamo states
with different kinetic and magnetic energies.
For still larger magnetic Reynolds numbers 
even a transition to chaos was predicted.

\begin{figure}[ht]
\begin{center}
\includegraphics[width=0.99\textwidth]{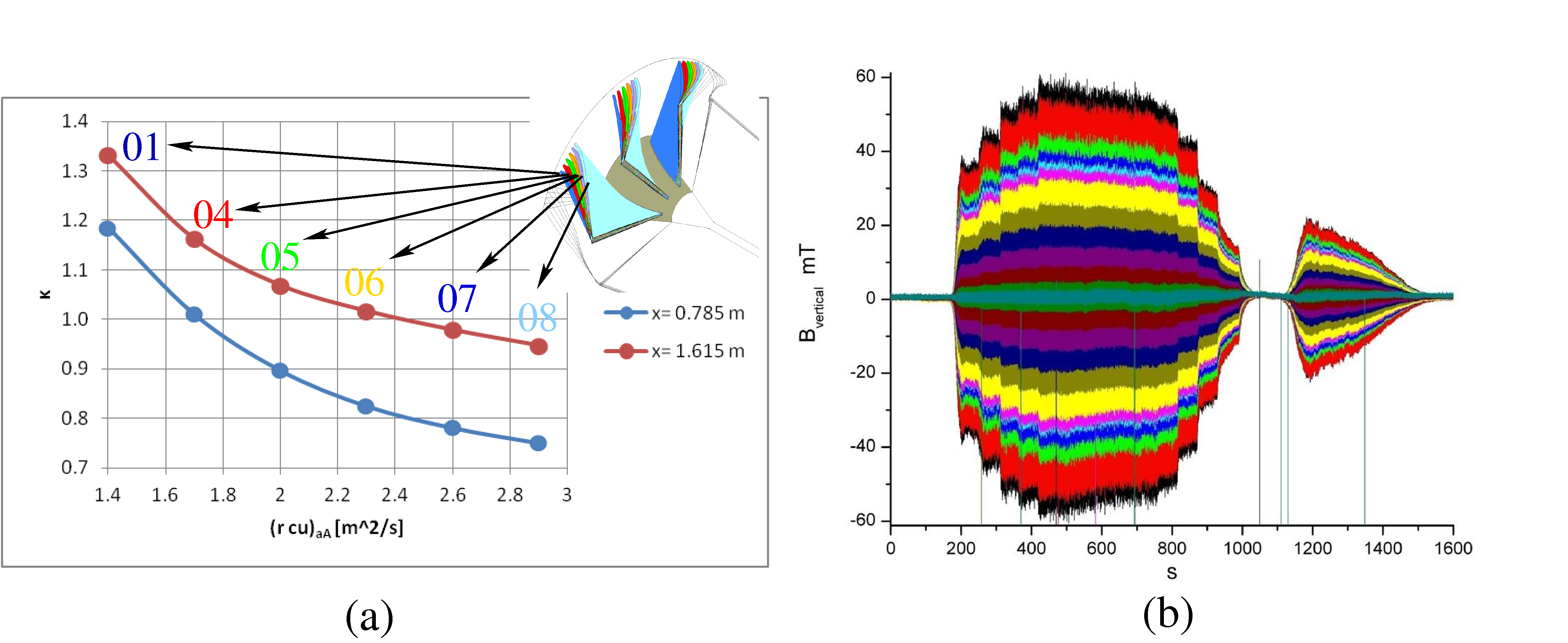}
\end{center}
\caption{Riga dynamo experiment. (a) Targeted  
''de-optimization'' of the flow profile behind the impeller by
choosing vanes with increasing pitch angles. The curves
show the ratio 
$\kappa=\sqrt{\overline{v}^2_{\rm z}/(2 \overline{v}^2_{\phi})}$ 
of mean axial to mean azimuthal velocity, in dependence on the 
swirl of the flow, for various vanes and 
two distances $x$ from the impeller. (b) Vertical magnetic
field measured at different radial position in the
first run after recommissioning (June 2016).}
\label{fig:riga}
\end{figure}

Based on these findings, one of the tasks of the 
A2 project was to figure out
how such scenarios could be realized in 
the Riga dynamo facility. We evaluated 
several technical provisions to  increase
the azimuthal velocity component beyond 
its optimal value, 
without completely destroying the flow 
structure that had been carefully 
optimized for the original Riga dynamo 
\cite{Stefani_1999}. A hydraulic 
analysis provided some feasible shapes and 
pitch angles
of the post-propeller vanes that indeed 
should produce the desired azimuthal velocities.
Figure  \ref{fig:riga}a shows the resulting
geometries and the corresponding velocity profile.
However, before implementing this new 
vanes' configuration we decided 
to first re-assembly the Riga dynamo in its old
form (with a stabilized second cylinder in order to
prevent buckling) and to validate
reproducibility of the former dynamo . 
Figure \ref{fig:riga}b shows the self-excited magnetic
field at different radial positions within the dynamo
as observed in the first experimental campaign 
after re-commissioning
in June 2016. More details can be found in \cite{Gailitis_2017}.

While the data are still under detailed analysis, 
we can conclude that the Riga dynamo experiment 
is now available for 
research into non-trivial back-reaction effects, 
and for testing measurement techniques for the 
DRESDYN precession dynamo.

\subsection{Numerical and experimental 
results on precession-driven flows in cylinders}

The largest installation in the framework 
of DRESDYN is a precession-driven 
liquid sodium experiment. Guided by early numerical 
estimations of the dynamo threshold of precession driven 
flows in cylinders 
\cite{Nore_2011,Stefani_2015,Cappanera_2016} and cubes
\cite{Krauze_2010}, a magnetic Reynolds 
number of ${\rm Rm \approx 700}$ was striven for. This ambitious 
number is achieved 
with a cylinder diameter of 2\,m rotating at 
 10\,Hz.

\begin{figure}[ht]
\begin{center}
\includegraphics[width=0.65\textwidth]{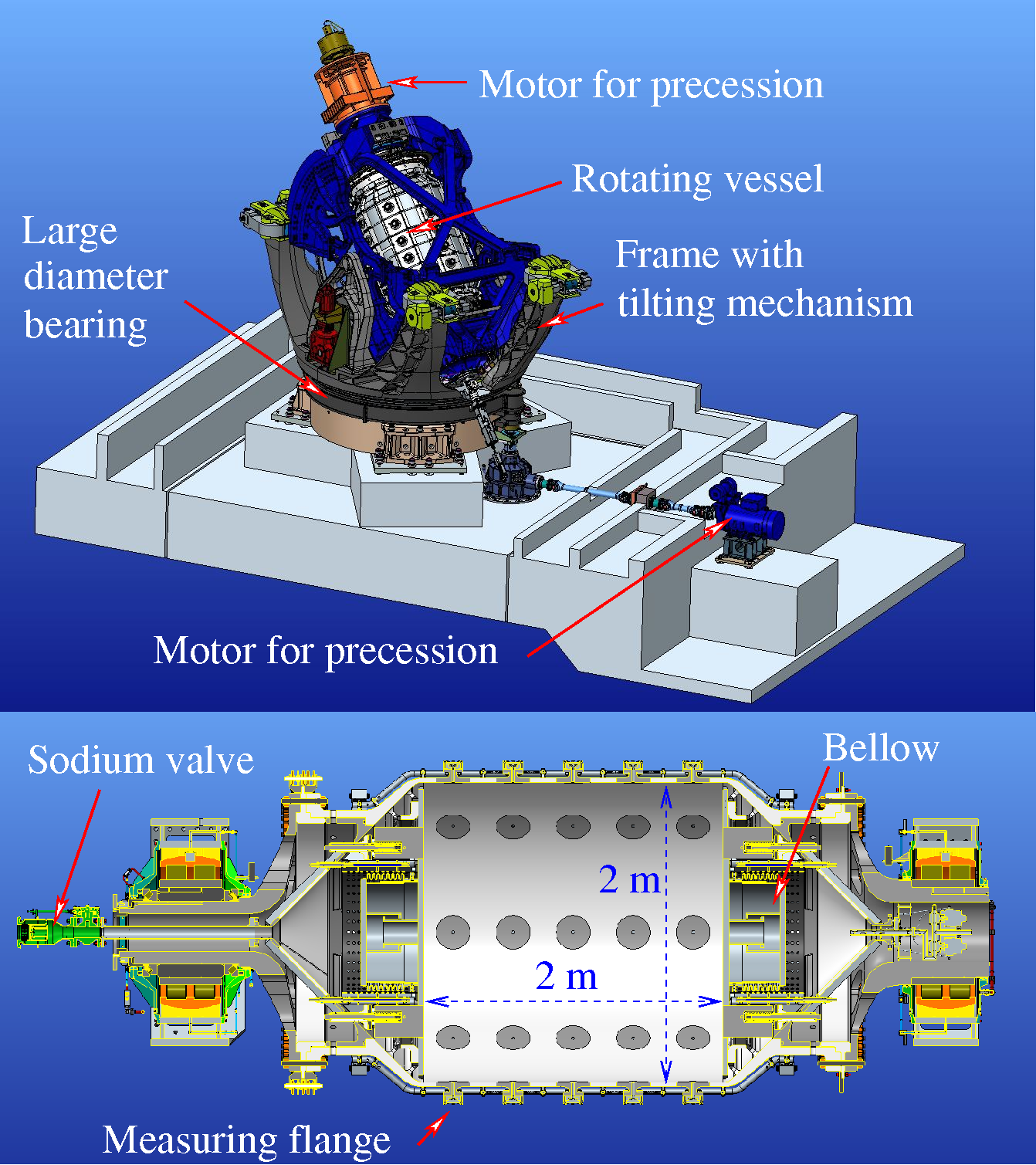}
\end{center}
\caption{Design of the precession-driven dynamo experiment. Figure 
courtesy SBS B\"uhnentechnik GmbH.}
\label{fig:dynamo}
\end{figure}

Design and construction of the precession dynamo
experiment (see Figure \ref{fig:dynamo}) is the 
responsibility of SBS B\"uhnentechnik GmbH. A  
detailed shape optimization of the rotating vessel 
turned out to be necessary 
in order to cope with the huge stresses that result
during precessing. The facility is presently 
under construction, and first pre-experiments with
water are expected for late 2018.

\begin{figure}[ht]
\begin{center}
\includegraphics[width=0.99\textwidth]{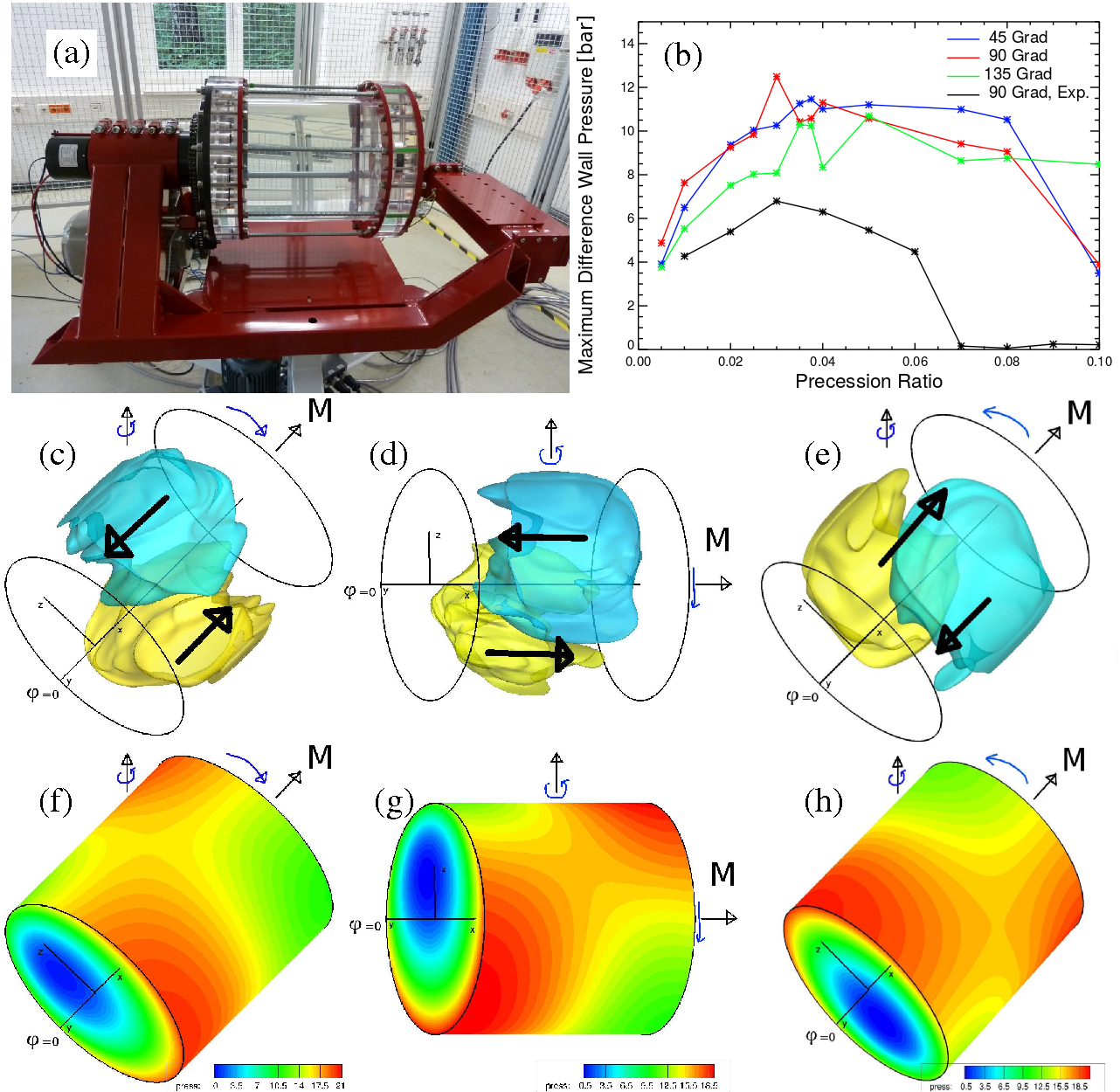}
\end{center}
\caption{Experimental and numerical preparations for the 
large precession experiment. (a) 1:6 down-scaled water experiment.
(b) Maximum pressure difference at the wall. The numerical curves,
computed at ${\rm Re}=6500$, are for the three different precession
angles 45$^{\circ}$ (blue), 90$^{\circ}$ (red), and 135$^{\circ}$ 
(green). The black curve gives the experimental data for 
90$^{\circ}$ measured at ${\rm Re}=1.6 \times 10^6$.
The middle row shows the numerically simulated axial velocity 
component for the 
angles 45$^{\circ}$ (c), 90$^{\circ}$ (d), and 135$^{\circ}$ (e),
the lower row (f,g,h) the corresponding pressure fields. }
\label{fig:precession}
\end{figure}

In preparing this large-scale experiment, 
a 1:6 scaled water experiment (see Figure \ref{fig:precession}a)
was built and utilized in order to gain
insight into the flow structure and the pressure 
field for varying Reynolds numbers and 
precession ratios (Poincar{\'e} numbers).
Parallel to that, the spectral element code
SEMTEX \cite{Blackburn_2004} was qualified 
and intensely used for precession-driven flows
in cylinders.
While such simulations are 
restricted to Reynolds numbers of about  ${\rm Re}=10^4$,
the large experiment  will reach approximately $10^8$.
The 1:6 water experiment reaches
a number of $1.6 \times 10^6$ when run at 10\,Hz, but can also
be slowed down to reach a value of $10^4$, 
thereby guaranteeing some overlap with numerical 
simulations.

A significant share of simulations and 
experiments was invested into estimating the
pressure field, a crucial input for 
the (static and dynamic) strength evaluation
of the vessel.  While 
pure rotation with 10\,Hz leads already 
to a centrifugal pressure of about 20\,bar,
it is the precession-driven pressure 
pulsation on the order of 10\,bar 
that makes the (dynamic) strength validation
so challenging.

Figure \ref{fig:precession} illustrates  the flow
structure (c,d,e) and the corresponding pressure 
field (f,g,h) computed at $\rm Re=6500$ and a Poincar\'e number
${\rm Po} \equiv \Omega_{\rm precession}/\Omega_{\rm rotation}=0.02$ 
for three angles between the rotation axis and the
precession axis 45$^{\circ}$, 
90$^{\circ}$, and 135$^{\circ}$.
The  maximum pressure pulsations in dependence on ${\rm Po}$ 
are shown in Figure \ref{fig:precession}b, together with 
the corresponding values resulting from the small water 
experiment (note that all values are up-scaled 
to the conditions of the large experiment). 
Basically, at low $\rm Po$  the 
flow is dominated by the first ($m = 1$) Kelvin mode 
which is mirrored
also by the increasing pressure pulsation.
Approximately at  ${\rm Po}=0.03$, higher azimuthal modes 
become relevant by drawing
more and more energy from the forced $m = 1$ 
Kelvin mode. The curve's kink at ${\rm Po}=0.07$ 
indicates a sudden transition of this
quasi-laminar regime 
to a turbulent regime, which is also confirmed
by the hysteretic behaviour of the motor power 
at this point
\cite{Herault_2015}.
The obvious difference between the simulated 
and experimental curves
is quite telling: transition to turbulence 
is strongly delayed in the (still highly viscous) 
numerical case which also shows significantly higher 
pressure pulsations. 

Apart from this rather global feature, our 
numerical simulation 
have also characterized the onset of higher $m$-modes
which usually are excited by triadic resonances 
with the forced $m=1$ Kelvin modes \cite{Giesecke_2015}.

\subsection{Precession-driven dynamos in cylinders and cubes}

Dynamo action of precession-driven flows is, 
unfortunately, a largely unsolved problem, despite 
a number of attempts to apply it to 
the dynamo of the Earth and other 
cosmic bodies 
\cite{Malkus_1968,Vanyo_2004,Tilgner_2005,Dwyer_2011,Fu_2012}.
The DRESDYN precession dynamo
experiment was partly motivated by 
the optimistic numerical
estimations obtained by Nore \cite{Nore_2011,Stefani_2015} 
for the cylinder, and
Krauze \cite{Krauze_2010} for a cube; both pointed 
consistently to
a critical ${\rm Rm}$ on the order of 700.

Unfortunately, this optimistic value is challenged 
when going to smaller values of the magnetic 
Prandtl number \cite{Giesecke_2015b}, i.e.\ to higher
Reynolds numbers.
Here, even the structure 
of the hydrodynamic flow, and its dependence 
on the precession ratio and the tilt angle,
is largely unknown. Recent simulations and experiments 
have  revealed the occurrence of 
higher $m$-modes \cite{Albrecht_2015,Giesecke_2015}
whose dynamo capabilities are still under scrutiny.
Even less is known about the turbulent flow structure when 
crossing the critical precession ratio.

Typically, direct numerical simulations for precessing flows
with correct no-slip boundary conditions work only until
${\rm Re} \approx 10^4$.
A promising way of going to much larger values
of ${\rm Re}$ is to replace no-slip boundary conditions
by stress-free conditions, thereby avoiding the 
need to resolve the viscous boundary layer. 
Such an approach was pursued by Goepfert and Tilgner 
 \cite{Goepfert_2016} who studied a precessing flow, and its
 dynamo action, in a cube. While this is by no means an 
 astrophysically relevant geometry, one may 
 naively expect the flow in a cube to resemble the flow in 
 the largest sphere enclosed by the cube, with some 
 dead water concentrated 
 in the corners. The main hope is that there 
 are some features common 
 to precessing flows in all geometries, such as the 
 appearance of triadic resonances, so that any geometry 
 is useful as a model system.
 
\begin{figure}[ht]
\begin{center}
\includegraphics[width=0.99\textwidth]{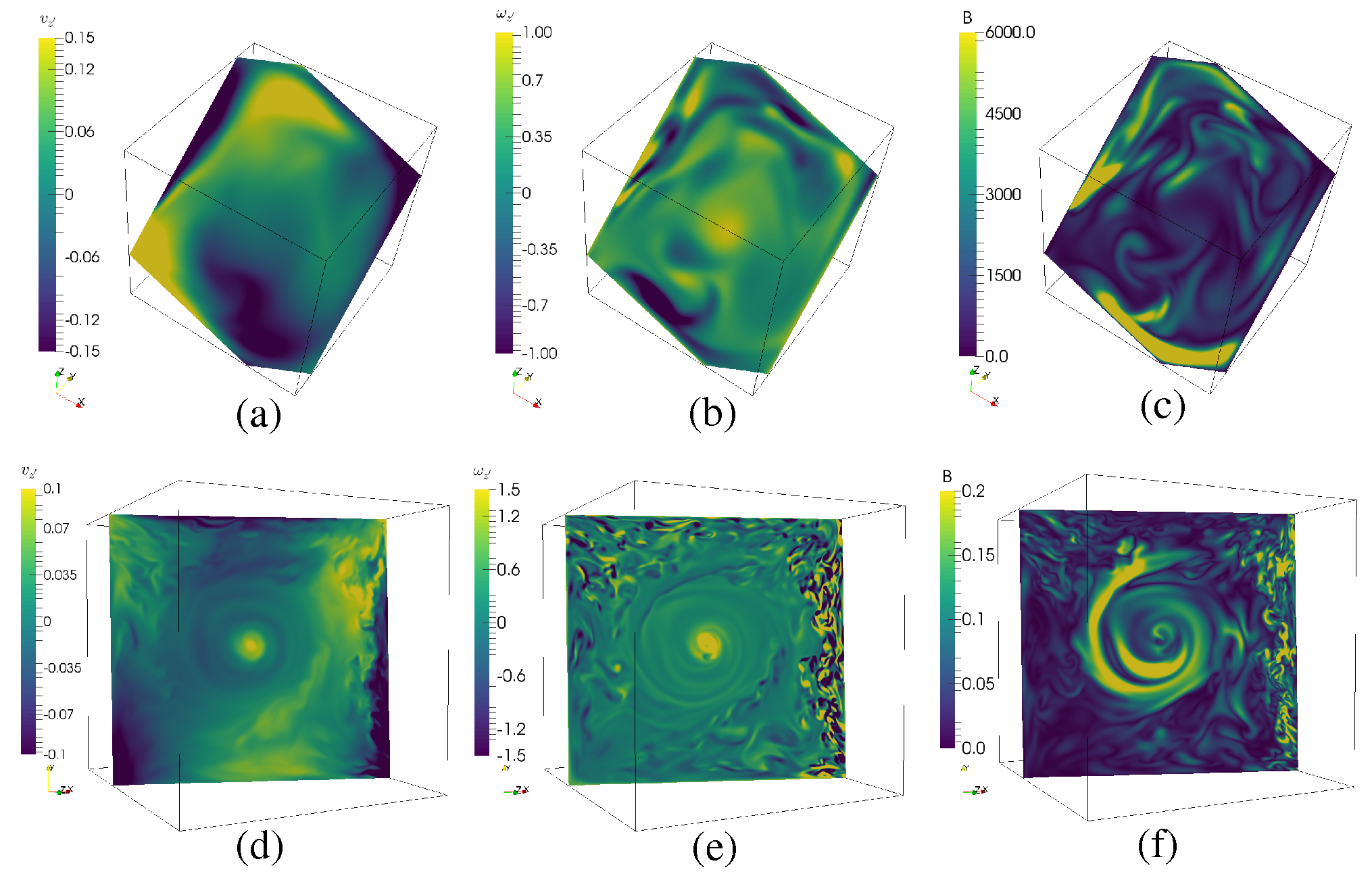}
\end{center}
\caption{Two different kinds of dynamos in a precessing cube. 
Upper row: $\mathrm{Re}=4 \times {10}^{3},{\rm{\overline{Po}   }}=-0.16$.
Lower row: $\mathrm{Re}={10}^{5},{\rm{\overline{Po}}}=-0.02$. (Note the
somewhat different definition of the Poincare number used here:
${\rm{\overline{Po}}}={\rm Po}/(1+{\rm Po} \cos{\alpha})$, with 
$\alpha$ denoting the precession angle).
Visualization of the velocity component
parallel to the rotation axis of the fluid 
(a,d), of the total vorticity (b,e), and of 
 the magnitude of the magnetic field (c,f)
in the plane perpendicular to this rotation 
axis.}
\label{fig:oliver}
\end{figure}

 The dynamo action of this flow depends quite sensitively,
 partly erratically,  
 on the Reynolds and the Poincare numbers. 
 Figure \ref{fig:oliver}
 illustrates the flow and the self-excited magnetic field
 for two different parameter sets. 

\section{Magnetically triggered flow instabilities}

We turn now to the field of magnetically triggered flow
instabilities, and report the main experimental
and theoretical results obtained within project A2.

\subsection{Experimental demonstration of AMRI,
and the large MRI/TI experiment}

After HMRI \cite{Stefani_2006b,Stefani_2009a} (and 
TI \cite{Seilmayer_2012}) had been experimentally 
demonstrated prior to the start of the LIMTECH 
Alliance, the PROMISE facility was qualified
for the experimental investigation
of AMRI \cite{Seilmayer_2014}.
Since AMRI needs a minimum central
current of about 10\,kA (given the material
parameters of GaInSn) the power supply for this current 
had to be enhanced to deliver 20\,kA.

\begin{figure}[ht]
\begin{center}
\includegraphics[width=0.99\textwidth]{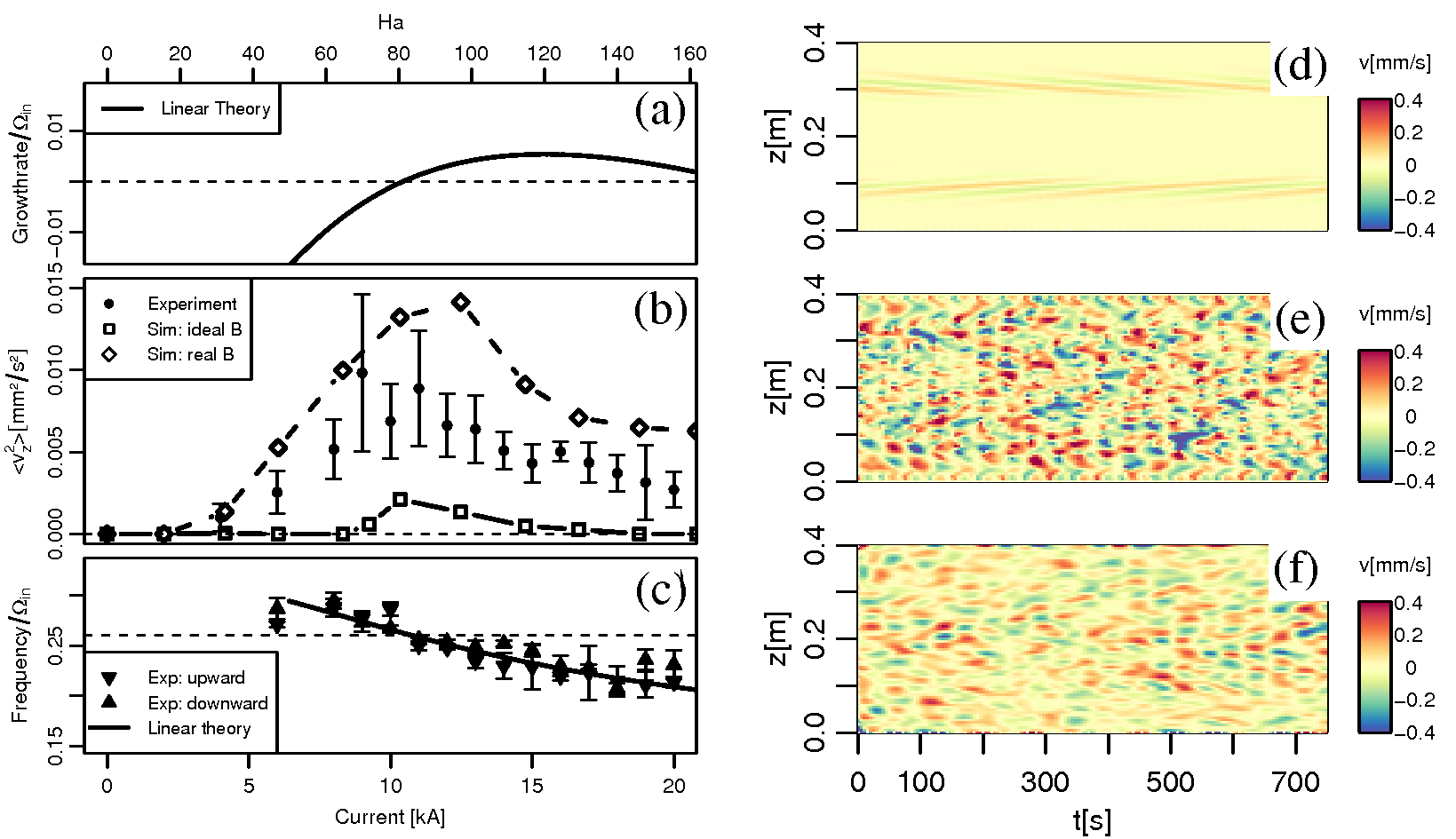}
\end{center}
\caption{Results of the AMRI experiment. Left: 
Dependence of various quantities on $\rm Ha$. 
(a) Numerically determined growth rate. (b) Mean 
squared velocity perturbation. (c) Angular drift frequency. 
In the frequency plot, ``upward'' and ``downward'' 
refer to the travel direction of the AMRI wave.
Right:
Velocity perturbation $v_z(m=1,z,t)$ for 
$\mu:=\Omega_{\rm out}/\Omega_{\rm in}=0.26$, $\rm Re=1480$,
and ${\rm Ha}=124$. (d) Simulation for ideal axisymmetric field. 
(e) Simulation for realistic field. (f) Experimental results. 
After \cite{Seilmayer_2014}.}
\label{fig:amri}
\end{figure}

Figure \ref{fig:amri} illustrates the dependence of various numerically and
and experimentally obtained quantities on the
central current (or the Hartmann number).
Figure \ref{fig:amri}a shows the growth rate
obtained with a 1D stability code, 
for ${\rm Re}=1480$, whose positive segment 
indicates the existence of 
AMRI in an interval approximately between 
10 and 22\,kA. Quite in correspondence with
that, the lowermost curve (''Sim: ideal B'') 
in Figure \ref{fig:amri}b 
shows the expected rms value of the
axial velocity, under the assumption of a 
purely azimuthal applied magnetic field. But here is a 
subtlety: The curve with the highest values (''Sim: real B'') 
in Figure \ref{fig:amri}b gives the 
corresponding numerical rms values
for the case that the real magnetic field 
of the experimental setting is 
assumed. This field deviates slightly from a pure $B_{\phi}$,
since the two leads from and to the power supply
form a single-winding coil. Although the resulting deviation
of the field is only in the region of a few 
per cent, the rms values are significantly enhanced,
and show also a broadened range for 
AMRI. Interestingly, the experimental  curve
 (''Experiment'') shows a very similar shape,
albeit with somewhat lower values.

The explanation of this phenomenon needs some peculiar
symmetry considerations: Let us start with an
infinitely long TC flow under the
influence of a purely azimuthal magnetic field.
As known from \cite{Hollerbach_2010}, this configuration
has no preference for $m=1$ or $m=-1$ modes, so
that both together would be expected to form
a standing wave. In a finite length TC cell,
the Ekman pumping at the lids produces slight
deviations from the perfect TC flow, which leads to
some preference for either $m=1$ or $m=-1$ 
modes in the upper and lower halves (see the simulation in
Figure \ref{fig:amri}d).
Adding now a second type of symmetry breaking, in form
of an imperfect applied magnetic field, the original
$m=\pm 1$ symmetry of the instability is to some extend
restored. In other words, the upward and downward 
travelling modes that would be expected to be
concentrated in the upper and lower part, now
interpenetrate each other (see simulation
in Figure \ref{fig:amri}e) and populate now
also the mid-height region, which effectively results
in significantly enhanced rms values.
This behaviour is indeed found in the experiment 
(Figure \ref{fig:amri}f).

While this simulated, and experimentally confirmed, 
effect of a double symmetry breaking on the AMRI is
interesting in its own
right, we have decided to improve the PROMISE facility 
in such a way 
that the azimuthal symmetry breaking is largely preserved.
This is realized by a new system of wiring of the 
central current, comprising now a ''pentagon'' of 5 
back-wires situated around the experiment. 
First experiments with this set-up show encouraging 
results, in particular transitions between AMRI and HMRI
when adding an axial field to the azimuthal one. 

\begin{figure}[ht]
\begin{center}
\includegraphics[width=0.6\textwidth]{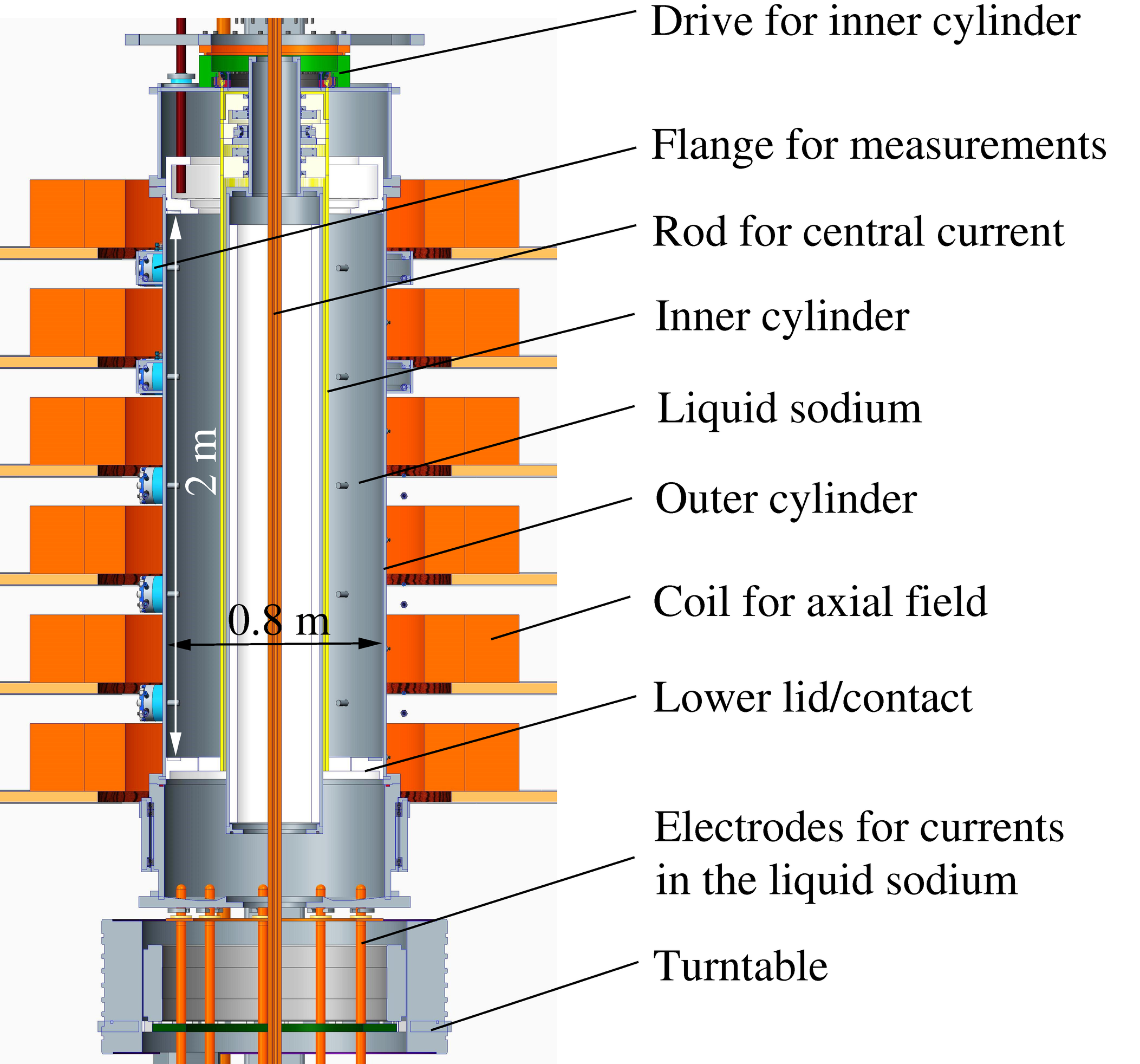}
\end{center}
\caption{Design of the large-scale TC experiment for investigations
of HMRI, AMRI, SMRI, TI and their combinations.}
\label{fig:mriti}
\end{figure}

The second large-scale sodium experiment to be set-up 
within the DRESDYN project 
aims at investigating  combinations of
different versions of the MRI and the current-driven
TI (see Figure \ref{fig:mriti}). Basically, the set-up is designed as 
a TC experiment with 2\,m
fluid height, an inner radius $r_{\rm in}=20$\,cm and an outer radius 
$r_{\rm out}=40$\,cm. 
Rotating the inner cylinder at up to 20\,Hz, we plan to reach 
${\rm Rm} \sim 40$, while the planned
axial magnetic field $B_z=120$\,mT will correspond 
to a Lundquist number $S:={\rm Pm}^{1/2}{\rm Ha} \sim 8$.
Both values are about twice the respective critical values 
for the onset of SMRI as
they were derived in \cite{Ruediger_2003}.

Below those critical values, 
we plan to investigate how HMRI approaches the limit 
of SMRI \cite{Hollerbach_2005,Kirillov_2010}. 
To this end, we will use a
strong central current, as it is already present 
in the PROMISE experiment \cite{Stefani_2009a,Seilmayer_2014}.
This insulated central current can be supplemented 
by another axial current guided
through the (rotating) liquid sodium, which will further 
allow to investigate
combinations of MRI and TI. Theoretical studies 
\cite{Kirillov_2013,Kirillov_2014,Priede_2015,Ruediger_2015}
have shown that even a slight addition of current through 
the liquid extends the range of application of the 
helical and azimuthal MRI to Keplerian flow profiles.

\subsection{Between HMRI, AMRI and TI: Some theoretical
results}

Shortly after the numerical revelation
of HMRI by Hollerbach and R\"udiger \cite{Hollerbach_2005},
Liu et al. \cite{Liu_2006} had 
derived -- in the framework of
a short-wavelength (or WKB) approximation -- 
two limits for the
negative and positive shear between 
which HMRI should cease to exist.
Expressed in terms of the 
Rossby number ${\rm Ro} \equiv r/(2 \Omega) \partial{\Omega}/\partial{r}$, 
the lower Liu limit (LLL) 
${\rm Ro}_{\rm LLL}=2(1-\sqrt{2}) \approx -0.828$ has
attracted a lot of interest \cite{Ruediger_2007,Priede_2009}, 
in particular
since it would prevent HMRI from working for
astrophysically interesting Keplerian flows characterized by
${\rm Ro}_{\rm Kepler}=-0.75$ . In contrast 
to this, the upper Liu limit 
${\rm Ro}_{\rm ULL}=2(1+\sqrt{2})\approx 4.828$ 
for positive shear flows 
has been largely ignored, although 
positive shear flows 
are indeed relevant in astrophysics, 
for example in a $\pm 30^{\circ}$ strip of the solar 
tachocline. We note in passing that 
positive shear flows were usually considered
perfectly stable (even under the action of vertical
magnetic fields), so that Deguchi's recent
discovery of a linear instability for
very large Reynolds numbers came as a big 
surprise \cite{Deguchi_2017}.

\begin{figure}[ht]
\begin{center}
\includegraphics[width=0.5\textwidth]{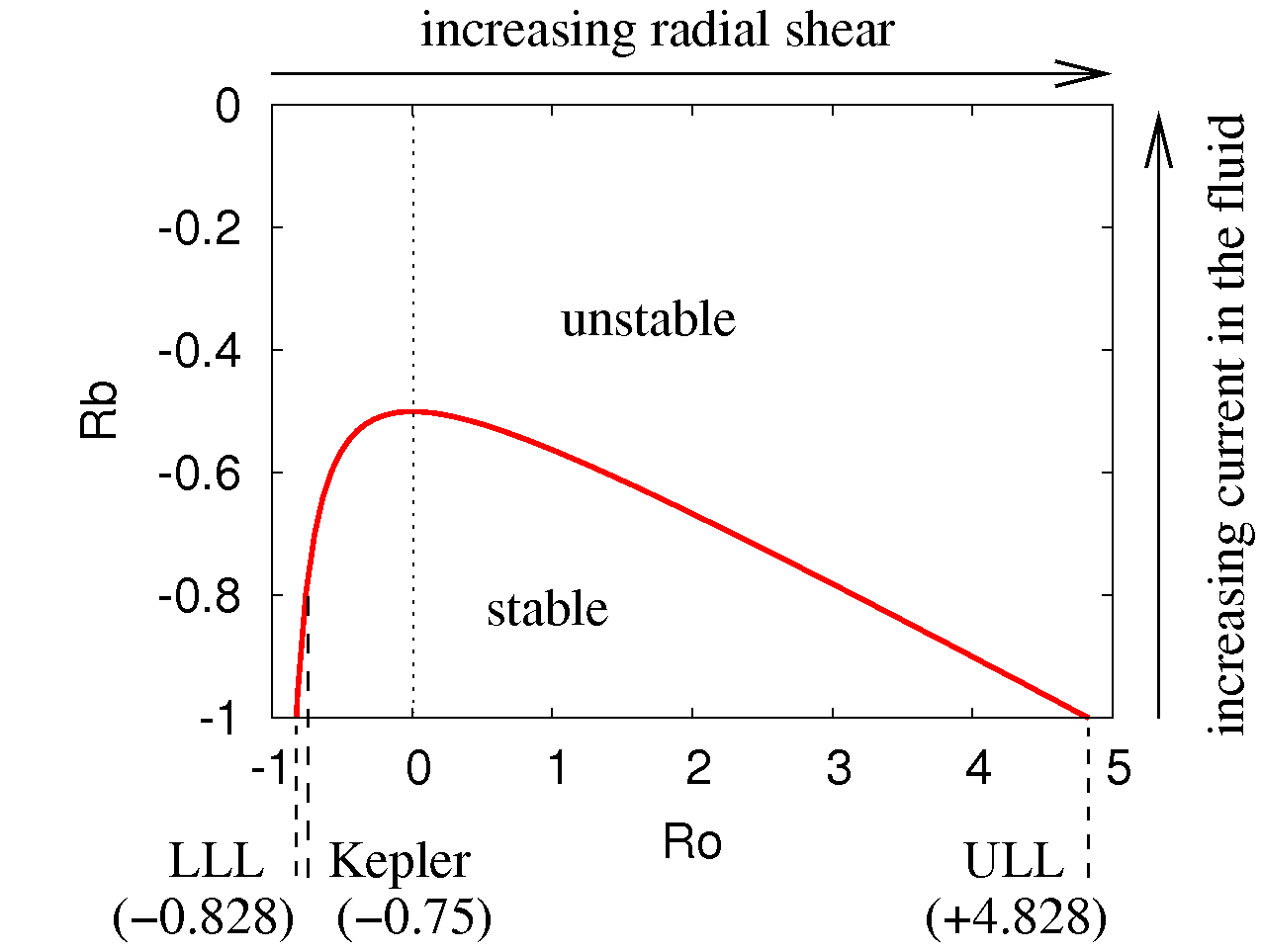}
\end{center}
\caption{Stability chart in the ${\rm Ro}-{\rm Rb}$ plane, 
for ${\rm Pm}=0$ and 
${\rm Ha}$ and ${\rm Re}$ tending to infinity. 
The Liu limits LLL and ULL apply only for 
${\rm Rb}=-1$, while for ${\rm Rb}>-1$ 
shallower shear profiles can as well 
be destabilized (including Kepler
rotation with ${\rm Ro}=-0.75$, 
starting at ${\rm Rb}=-0.78125$). The dotted line separates flows
with negative shear (to the left) and positive shear (to the right).}
\label{fig:oleg}
\end{figure}

A first interesting result regarding the two Liu limits
was published in  \cite{Kirillov_2013}.
This work dealt with the question how HMRI (and AMRI)
would be modified if the central current, 
which generates the azimuthal field component, was
gradually complemented by some second axial 
current through the liquid. The respective weight 
of the two azimuthal 
field parts can be 
quantified by the magnetic Rossby number ${\rm Rb} \equiv r^2/(2 B_{\phi}) 
\partial(r^{-1} B_{\phi})/\partial{r}$
which is constructed as a counterpart of ${\rm Ro}$. 
A pure current-free field (as assumed  
for HMRI and AMRI \cite{Kirillov_2012})
gives ${\rm Rb}=-1$, while a pure current in 
the fluid (as for TI) corresponds to ${\rm Rb}=0$.
In \cite{Kirillov_2013}, and more detailed in
\cite{Kirillov_2014}, it was shown that 
the LLL and the ULL are just the endpoints of one
common stability curve (Figure \ref{fig:oleg})
in the ${\rm Ro}-{\rm Rb}$ plane 
which acquires the surprisingly simple 
analytical form 
\begin{equation}
{\rm Rb}=-\frac{1}{8}\frac{({\rm Ro}+2)^2}{{\rm Ro}+1},
\label{rel}
\end{equation}
when both ${\rm Re}$ and ${\rm Ha}$ 
are sent to infinity and either the ratio of azimuthal 
to axial field (for HMRI) or the shape of the 
perturbation (for AMRI) is optimized.
The ''metamorphosis'' of the helical MRI when
changing the ratio of the axial currents within 
the fluid and on the axis 
were also studied with a 1D stability 
code \cite{Priede_2015}.

The treatment of problems with variable ${\rm Rb}$
led also to the reconsideration of an old
problem of magnetohydrodynamics which is 
known as Chandrasekhar's theorem \cite{Chandra_1956}.
In the framework of ideal MHD this theorem states
the stability of rotating flows of any radial
dependence under the influence of an
azimuthal magnetic field whose corresponding Alfv{\'e}n
velocity has the same amplitude and radial
dependence as the rotation. 
Both in the WKB framework \cite{Kirillov_2014} and
with a 1D-stability code \cite{Ruediger_2015} 
we showed that these Chandrasekhar solutions can 
be destabilized in non-ideal MHD.

We also mention the paper by Priede \cite{Priede_2017}
that, treating a simplified pinch-type instability in a 
semi-infinite planar sheet of an inviscid incompressible 
liquid with a straight rigid edge, helped to clarify 
some differences between the instability mechanisms 
in highly resistive and
well conducting fluids, resulting  
in different development times: 
magnetic response time for the former,
and the much shorter Alfv{\'e}n time for the latter.

\subsection{Super-AMRI}

\begin{figure}[ht]
\begin{center}
\includegraphics[width=0.99\textwidth]{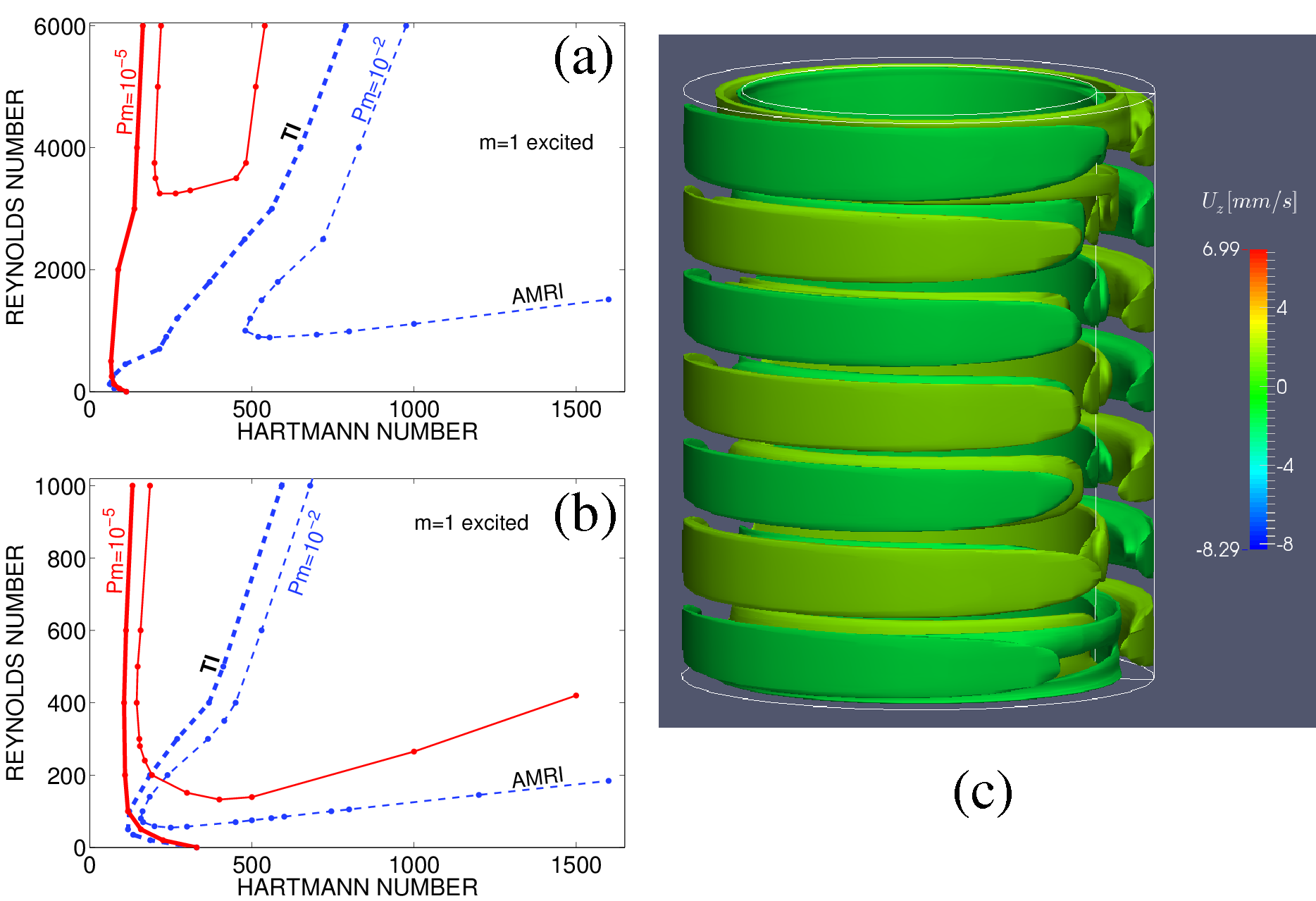}
\end{center}
\caption{Azimuthal MRI for flows with positive shear (Super-AMRI): 
(a) Stability maps in the Hartmann-Reynolds plane 
for resting inner cylinder with 
$r_{\rm in}/r_{\rm out}=0.75$  and two magnetic Prandtl numbers
$10^{-5}$ (red) and $10^{-2}$ (blue), without (AMRI) and 
with (TI) current in the 
fluid. (b) The same for $r_{\rm in}/r_{\rm out}=0.9$. 
(c) Simulated axial velocity 
component for $r_{\rm in}/r_{\rm out}=0.75$.}
\label{fig:super}
\end{figure}

Within the project, we have proved the existence of the 
upper Liu limit also for purely azimuthal magnetic fields,
both in WKB approximation \cite{Stefani_2015b} 
and with a 1D stability code \cite{Ruediger_2016,Gellert_2016}. 
The seed grant ''Super-AMRI'' was dedicated to
find optimal parameters  for 
a liquid metal TC experiment to show this ''Super-AMRI'', as
we call it now. 
The main problem here is that such a TC experiment
would need a rather thin gap in order to realize the
enormous positive shear of ${\rm Ro}>4.828$, which 
leads to a very large critical value of the 
central current.
For a liquid sodium experiment, both 
WKB \cite{Stefani_2015b} 
and 1D simulations with insulating boundary 
conditions \cite{Gellert_2016} 
point to a critical current of about 80 kA,
while a much lower (and experimentally more feasible) 
value of some 20 kA results for ideally conducting 
cylinder walls \cite{Gellert_2016}. 
Figure \ref{fig:super} shows the stability maps 
in the Hartmann-Reynolds plane 
for $r_{\rm in}/r_{\rm out}=0.75$ (a) and $r_{\rm in}/r_{\rm out}=0.9$ (b) and 
two magnetic Prandtl numbers
$10^{-5}$ (red) and $10^{-2}$ (blue), 
without (AMRI) and with (TI) current in the 
fluid, together with the simulated 
axial velocity 
component (c) for the case $r_{\rm in}/r_{\rm out}=0.75$.
The technically feasible  
combination of using sodium and copper 
walls is expected to lead to a critical
current 
between 20\,kA and 80\,kA, but a detailed simulation 
for that case
is still in progress.

Hence, it might  be worthwhile to come back
to the idea of ''Super-HMRI'', for which a
1D simulation is also pending. With view 
on the significant differences in the 
axial currents that are needed for the
experimental observation of the negative-shear 
versions of HMRI (4\,kA) and AMRI (10\,kA),
we expect a similar reduction 
in the positive shear case that could
reduce the technical efforts significantly.

\subsection{Linking dissipation-induced instabilities and non-modal growth}
While the two Liu limits for HMRI have been
known for more than a decade \cite{Liu_2006},
the physical reason behind them was never brought into 
question.
In a recent paper \cite{Mama_2016}, we have 
revealed a link between the modal growth {\it rate} 
of HMRI (from which the Liu limits follow) 
and the non-modal growth {\it factor}
of the underlying purely hydrodynamical problem,
which sheds some new light on the physical 
essence of both.

Non-modal (or transient) growth is typical for 
the time evolution of dynamical systems governed by 
non-normal operators. Due to the 
non-orthogonality of their eigenfunctions,
an appropriately chosen initial state can
experience a transient growth of its amplitude
even if the (two or more) individual eigenfunctions
of which it is composed have negative modal 
growth rates. Non-modal instabilities are known to
play a key role in explaining the onset of 
turbulence in pipe flows \cite{Trefethen_1999}. 

\begin{figure}[ht]
\begin{center}
\includegraphics[width=0.5\textwidth]{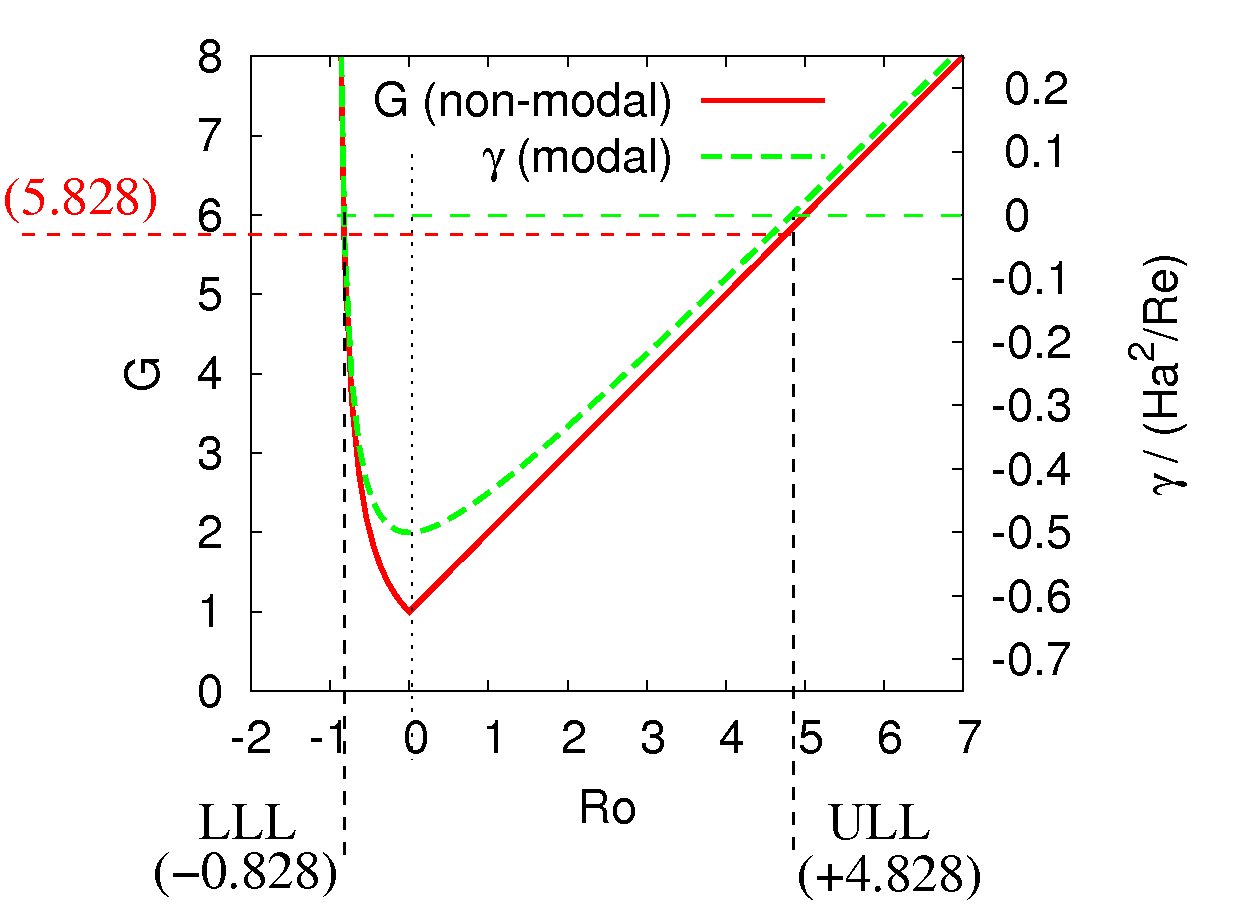}
\end{center}
\caption{Non-modal growth factor $G$ for purely hydrodynamic
rotating flows (red curve), and normalized growth rate $\gamma$ of 
helical MRI (green curve),  in dependence on the Rossby number
$\rm Ro$. The two quantities are connected by Equation (2) and 
match asymptotically according to $\gamma \approx {\rm Ha}^2/{\rm Re}(G/8 -3/4)$.}
\label{fig:mama}
\end{figure}

As shown by Afshordy et al. \cite{Afshordi_2005}, 
the non-modal growth factor for
axisymmetric perturbations
of purely hydrodynamic rotating flows can be 
expressed by the surprisingly simple
equation $G=(1+{\rm Ro})^{sgn({\rm Ro})}$ which is 
illustrated by the red curve in Figure \ref{fig:mama}.
Comparing this with the equation for the modal 
growth rate $\gamma$ (green curve in Figure \ref{fig:mama}) 
of HMRI \cite{Kirillov_2014}, 
we find the following link \cite{Mama_2016}:
\begin{eqnarray} 
\gamma=\frac{{\rm Ha}^2}{\rm Re} \left[ \frac{({\rm Ro}+2)^2}{8 ({\rm Ro}+1)}-1  \right]
=\frac{{\rm Ha}^2}{\rm Re} \left[ \frac{(G+1)^2}{8 G}-1  \right] \; .
\end{eqnarray}
For large $|Ro|$ this leads to a simple linear 
relation of
$\gamma$ and $G$: $\gamma \approx {\rm Ha}^2/{\rm Re}(G/8 -3/4)$.
Figure \ref{fig:mama} illustrates 
this asymptotic connection. At the two Liu limits of HMRI,
${\rm Ro}=2(1 \pm \sqrt{2})$, $G$ acquires the 
same value $G=1+2(1+\sqrt{2})\approx 5.828$.

Keeping in mind that HMRI is a (double-)diffusive 
instability, we learn that it is inherently based on 
the non-modal growth of the underlying purely hydrodynamic
flow. For a linear instability to become 
effective, it requires some dissipation to allow a
coupling between different components of the eigenfunctions
(actually between meridional and azimuthal flow
perturbations, as shown in \cite{Priede_2007}).
Once this coupling is established, 
the growth rate of the linear HMRI 
becomes (nearly) proportional to the
non-modal growth factor of the
underlying hydrodynamic flow.

It remains to be seen whether similar links exist also
in other fields of hydrodynamics. This is not 
obvious, as we have already experienced when analyzing 
the corresponding connection for AMRI \cite{Mama_2017} 
that is more subtle due to the shear-induced time-variation
of the wavenumber of the $m=1$ mode.

\section{Conclusions}
Two large-scale liquid sodium experiments,
on precession and on MRI/TI,
are presently under construction at HZDR.
For both of them, the project A2 of the
LIMTECH Alliance has made theoretical and
experimental contributions.
The project has further supported
re-commissioning of the Riga dynamo experiment 
which is now ready for studying
new interesting back-reaction effects as well as 
for testing measurement techniques for the DRESDYN 
experiments. 
A particular achievement of the project 
is a deepened insight into the 
diffusive instabilities of
rotating flows under the action 
of magnetic fields. As an outcome of those activities,
the starting grant on ''Super-AMRI'' has delivered
first parameter estimates 
for a new liquid sodium experiment
on the magnetic destabilization of positive shear flows. 
Another result concerns the 
rigorous mathematical link between the non-normal growth factor
of purely hydrodynamic rotating flows 
with the normal growth rate of the
HMRI which gives -- to the best of our knowledge -- 
the first link between
non-normal growth and 
diffusive (or dissipation-induced) instabilities.
We should also mention here the recent astrophysical 
application of AMRI for explaining the 
angular momentum redistribution of
post-main sequence low-mass stars \cite{Spada_2016}.

Many problems related to the TI liquid metals 
were investigated in close collaboration with the 
project B3 which was dedicated to applied problems
of liquid metal batteries \cite{Weber_2013,Weber_2014,Weber_2017}. 
As one of the potentially 
dangerous instabilities, which would occur even if 
electrovortex flows and sloshing instabilities
were completely suppressed, the TI was numerically
treated with an OpenFOAM code enhanced by a Poisson solver for the
electric potential and Biot-Savart's law for the 
induced magnetic field
\cite{Weber_2013}. 
With this code it was possible to test various 
provisions for suppressing the TI and to understand
its saturation mechanism \cite{Weber_2014}. 
An interesting by-product 
of these simulations was the detection of helicity 
oscillations \cite{Weber_2015} for  
small magnetic Prandtl numbers. Re-applying this result 
(obtained for liquid metal batteries) to a simple 
model of a Tayler-Spruit solar dynamo,
we identified a mechanism that could empower the 
weak tidal forces of planets to synchronize the 
Hale cycle of the solar magnetic field 
\cite{Stefani_2016,Stefani_2017}.
It is here where the synergies between basic research 
and liquid metal technologies, which were fostered by 
the LIMTECH Alliance, proved particularly useful.

\ack
This work was supported by Helmholtz-Gemeinschaft 
in frame of the LIMTECH Alliance. Close cooperation 
with Tom Weier and Norbert Weber (project B3) 
on various aspects of the  Tayler instability 
is gratefully acknowledged.

\end{document}